\documentclass[11pt]{article}
\usepackage{latexsym}
\usepackage{graphicx}
\usepackage{amsfonts,amssymb}

\begin{document}

\title{A new mathematical representation of Game Theory II}
\author{Jinshan Wu \\
Department of Physics, Simon Fraser University, Burnaby, B.C. Canada, V5A 1S6%
}
\maketitle

\begin{abstract}
In another paper with the same name\cite{frame}, we proposed a new
representation of Game Theory, but most results are given by
specific examples and argument. In this paper, we try to prove the
conclusions as far as we can, including a proof of equivalence
between the new representation and the traditional Game Theory,
and a proof of Classical Nash Theorem in the new representation.
And it also gives manipulation definition of quantum game and a
proof of the equivalence between this definition and the general
abstract representation. A Quantum Nash Proposition is proposed
but without a general proof. Then, some comparison between Nash
Equilibrium (NE) and the pseudo-dynamical equilibrium (PDE) is
discussed. At last, we investigate the possibility that whether
such representation leads to truly Quantum Game, and whether such
a new representation is helpful to Classical Game, as an answer to
the questions in \cite{enk}. Some discussion on
continuous-strategy games are also included.
\end{abstract}

Key Words: Game Theory, Nash Equilibrium, Quantum Game, Continuous
Strategy.

Pacs: 02.50.Le, 03.67.Lx

\section{Introduction}
\label{intro}

In \cite{frame}, we have pointed out that all ideas and concepts
in traditional framework of Game Theory can be translated into our
new representation, by some specific examples of discrete-strategy
games. Now by giving some proof, especially a proof of Nash
Theorem, we wish to confirm that this new representation can
express every idea in traditional framework of classical game.

On the other hand, for quantum games, as pointed out in
\cite{enk}, two questions should be answered when a quantum
framework or a quantized version of classical game is discussed to
compare with the corresponding classical game. The first one is
whether the new approach is helpful to solve the classical one,
the real original classical game, not the quantized version. The
second one is whether the quantized version is a truly quantum
mechanics problem with independent meaning other than the
corresponding classical one. In this paper, we try to give a
positive answer to those two questions, although frankly speaking,
our new representation of game theory\cite{frame} is not exactly
the same with quantum games in \cite{meyer,jens}.

In the next section ($\S\ref{form}$), we shortly review the
structure of our new representation and results when applied onto
discrete-strategy game. Then classical Nash Theorem in the new
representation is proved in section $\S\ref{nash}$ and compared
with our Pseudo-dynamical Equilibrium (PDE)($\S\ref{compare}$). In
$\S\ref{quantum}$, Quantum Game is defined from the manipulation
level starting from quantum object and quantum operators. Also NE
and Nash Theorem for quantum game is proposed there, but not
proved. The last part ($\S\ref{remark}$) is devoted to answer the
two questions mentioned above, while a short summary of the
results in this paper is also included.

\section{The new representation of classical game}
\label{form}

Traditionally, a classical static non-cooperative game is defined
as
\begin{equation}
\Gamma^{c} = \left(\prod_{i=1}^{N}\otimes S_{i},
\left\{G^{i}\right\}\right), \label{oldgame}
\end{equation}
in which $S_{i}$ is the $L_{i}$-element strategy set of player
$i$, and $G^{i}$ is a mapping from $\prod_{i=1}^{N}\otimes S_{i}$
to $\mathbb{R}$. A state of player $i$ can be a mixture strategy
as
\begin{equation}
\vec{P}^{i} = \left(p^{1}_{1}, \cdots, p^{1}_{\mu}, \cdots,
p^{1}_{L_{i}} \right)^{T},
\end{equation}
in which $p^{i}_{\mu}$ is the probability that player $i$ choose
strategy $\mu$ from the set $S_{i}$. The payoff value of player
$i$ is
\begin{equation}
E^{i}\left(\vec{P}^{1}, \cdots, \vec{P}^{N}\right) =
\sum_{s^{1}_{\mu},\cdots,s^{N}_{\nu}}G^{i}_{s^{1}_{\mu}\cdots
s^{N}_{\nu}}p^{1}_{\mu}\cdots p^{N}_{\nu}.
\label{oldpayoff}
\end{equation}
So for state vector $\vec{P}^{i}$, $\left\{G^{i}\right\}$ is a set
of $\left(0,N\right)$-tensor. A Nash Equilibrium state
$\vec{P}^{1}_{eq}, \cdots, \vec{P}^{N}_{eq}$ is defined that
\begin{equation}
E^{i}\left(\vec{P}^{1}_{eq}, \cdots, \vec{P}^{i}_{eq}, \cdots,
\vec{P}^{N}_{eq}\right) \geq E^{i}\left(\vec{P}^{1}_{eq}, \cdots,
\vec{P}^{i}, \cdots, \vec{P}^{N}_{eq}\right), \forall i, \forall
\vec{P}^{i}. \label{oldne}
\end{equation}
For a continuous-strategy game, $\vec{P}^{i}$, the state vector of
player $i$, will be a probability distribution function on the
continuous strategy set, and all summations turn into integral.

In our new representation, we defined base vector and inner
product in strategy sets so as to form them as Hilbert space, and
then the system state space is the direct product space of all
single-player spaces. The base vectors of player $i$'s strategy
space are chosen as all the pure strategies, and denoted as
$\left|s^{i}_{\mu}\right>$, the inner product is defined as
\begin{equation}
\left<s^{i}_{\mu}\left|s^{i}_{\nu}\right>\right. = \delta{\mu\nu}.
\end{equation}
So base vectors of system state space are
\begin{equation}
\left|S\right> =
\left|s^{1}_{\mu}\right>\cdots\left|s^{N}_{\nu}\right> \triangleq
\left|s^{1}_{\mu},\cdots,s^{N}_{\nu}\right> .
\end{equation}
Then, a state of player $i$ expressed in density matrix form is
\begin{equation}
\rho^{i} =\sum_{\mu} p^{i}_{\mu}\left|\mu\right>\left<\mu\right|.
\end{equation}
So a system state is
\begin{equation}
\rho^{S} \triangleq \prod_{i}^{N}\rho^{i} = \sum_{S}
p^{1}_{\mu}\cdots p^{N}_{\nu}\left|S\right>\left<S\right|.
\end{equation}
According to Quantum Mechanics, the expectation value should be
calculated by $E = Tr\left(\rho^{S}H\right)$, in which $H$ is a
quantum operator defined on system state space. We want to keep
the same mathematical form, so the payoff value of player $i$ must
be
\begin{equation}
E^{i} = Tr\left(\rho^{S}H^{i}\right).
\label{newpayoff}
\end{equation}
Now we need to define a payoff matrix $H^{i}$ to guarantee that
the payoff from equ(\ref{newpayoff}) are equivalent with the
payoff from equ(\ref{oldpayoff}). Under such requirement, we find
\begin{equation}
\begin{array}{ccc}
H^{i}_{SS^{'}} = G^{i}_{S}\delta_{SS^{'}} & \mbox{or} & H^{i} =
\sum_{SS^{'}}
G^{i}_{S}\delta_{SS^{'}}\left|S\left>\right<S^{'}\right|
\end{array},
\label{newpayoffmatrix}
\end{equation}
in which
\begin{equation}
G^{i}_{S} = G^{i}_{s^{1}_{\mu}\cdots s^{N}_{\nu}}.
\end{equation}
It's easy to check the equivalence,
\begin{equation}
\begin{array}{ccc}
E^{i} = Tr\left(\rho^{S}H^{i}\right) & = & \sum_{S^{'}}
\left<S^{'}\left|\rho^{S}H^{i}\right|S^{'}\right>
\\& = & \sum_{S^{'},S}\left<S^{'}\left|\rho^{S}\left|S\left>\right<S\right|H^{i}\right|S^{'}\right>
\\& = & \sum_{S}\left<S\left|\rho^{S}\right|S\right>H^{i}_{SS}
\\& = & \sum_{S}\rho^{S}_{SS}H^{i}_{SS}
\\& = & \sum_{S}p^{1}_{\mu}\cdots p^{N}_{\nu}G^{i}_{s^{1}_{\mu}\cdots
s^{N}_{\nu}}
\end{array}.
\label{proof1}
\end{equation}
Now in our new representation, a game is redefined as
\begin{equation}
\Gamma^{c,new} = \left(\prod_{i=1}^{N}\otimes S_{i},
\left\{H^{i}\right\}\right). \label{newgame}
\end{equation}
So we have

\noindent {\bf{Theorem I}} Classical game $\Gamma^{c,new}$ is
equivalent with game $\Gamma^{c}$.

NE is redefined as $\rho^{S}_{eq}$ that
\begin{equation}
E^{i}\left(\rho^{1}_{eq}\cdots\rho^{i}_{eq}\cdots\rho^{N}_{eq}\right)
\geq
E^{i}\left(\rho^{1}_{eq}\cdots\rho^{i}\cdots\rho^{N}_{eq}\right),
\forall i, \forall \rho^{i}. \label{newne}
\end{equation}
A reduced payoff matrix, which means the payoff matrix when all
other players' states are fixed, is defined as
\begin{equation}
H^{i}_{R} =
Tr_{-i}\left(\rho^{1}\cdots\rho^{i-1}\rho^{i+1}\cdots\rho^{N}H^{i}\right),
\end{equation}
in which $Tr_{-i}\left(\cdot\right)$ means to do the trace in the
space except the one of player $i$. If a trace in player $i$'s
space is needed, we denote it as $Tr^{i}\left(\cdot\right)$.
Payoff value can also be calculated by the reduced matrix as
\begin{equation}
E^{i} = Tr^{i}\left(\rho^{i}H^{i}_{R}\right).
\end{equation}

\section{Continuous strategy space}
\label{continue}

For a continuous-strategy game, we need to replace summation
($\sum_{\mu}$) with integral ($\int d\mu$) and to replace
probability $p^{i}_{\mu}$ with probability density function
$p^{i}\left(\mu\right)$. And then, the inner product will be
\begin{equation}
\left<\mu^{i}\left|\nu^{i}\right>\right. =
\delta\left(\mu^{i}-\nu^{i}\right).
\end{equation}
In fact, before we define this, we have to claim what's $d\mu$,
the measure of $\mu$. But here, let's say our continuous
strategies are something like price, so that a nature measure is
predefined. Another problem is the normalization condition. As
Quantum Mechanics, we have to deal with the term like
$\left<\mu\left|\mu\right>\right.$. In Quantum Mechanics, it
doesn't matter if wave function is used as a description of state.
On the other hand, from equ(\ref{proof1}), we know that because
$H^{i}$ has only diagonal terms, only diagonal part of $\rho^{i}$
will effect the payoff. This means that if another density matrix
but with the same diagonal part is used to describe state, it will
give the same payoff value. In Quantum Mechanics viewpoint, a pure
state described by a wave function can be used to replace the
density matrix of mixture state if
\begin{equation}
\rho^{i}_{ss} = \phi^{*}(s)\phi(s).
\label{wavefunction}
\end{equation}
Then, in a continuous-strategy game, a state of player $i$ is
$\left|\phi\right> = \int d\mu \phi(\mu)\left|\mu\right>$, or
density matrix form,
\begin{equation}
\rho^{i} = \int\int d\mu d\nu
\phi^{*}(\nu)\phi(\mu)\left|\mu\right>\left<\nu\right|.
\end{equation}
But only diagonal term $\phi^{*}(\mu)\phi(\mu)$ will effect the
payoff. The normalization condition $\int d\mu
\left<\mu\left|\rho^{i}\right|\mu\right> = 1$ gives
\begin{equation}
\int d\mu \phi^{*}(\mu)\phi(\mu) =1.
\end{equation}
The traditional payoff function and the relation with payoff value
is
\begin{equation}
E^{i}= \int dS p^{1}\left(s^{1}\right)\cdots
p^{N}\left(s^{N}\right)G^{i}\left(s^{1}, \cdots, s^{N}\right),
\label{oldc}
\end{equation}
in which, according to eu(\ref{wavefunction}), $p^{i}(\mu) =\left|
\phi^{i}\left(\mu\right)\right|^{2}$. And in our new
representation,
\begin{equation}
E^{i}= \int dS
\left<S\right|\rho^{1}\cdots\rho^{N}H^{i}\left|S\right>,
\label{newc}
\end{equation}
in which
\begin{equation}
H^{i}= \int\int dSdS^{'}
\left|S\left>\right<S^{'}\right|G^{i}\left(S\right)\delta(S-S^{'}).
\end{equation}
It's easy to prove equ(\ref{oldc}) and equ(\ref{newc}) give the
same payoff value. Although here all formulas are derived in
continuous strategy space form, for discrete strategy games, pure
state with condition $\left|\phi^{i}_{\mu}\right|^{2} =
p^{i}_{\mu}$ can also be used to replace density matrix form of
mixture state and give the same payoff.

\section{Proof of Nash Theorem}
\label{nash}

Nash Theorem proves the existence of NE. For a game defined by
equ(\ref{oldgame}), equilibrium states defined by equ(\ref{oldne})
always exist. Now in our new representation, Nash Theorem is
reexpressed as

\noindent{\bf{Theorem II}} For a game defined by
equ(\ref{newgame}), equilibrium states defined by equ(\ref{newne})
always exist.

\noindent{\bf{Proof}} Just following the idea of Nash's proof,
first, we define a mapping, and prove the existence of the fixed
points of this mapping. Then we will prove the fixed points are
NE.

A mapping is defined as
\begin{equation}
\left(\rho^{1,'}, \cdots, \rho^{N,'}\right) =
\mathcal{T}\left(\rho^{1}, \cdots, \rho^{N}\right),
\end{equation}
in which,
\begin{equation}
\rho^{i,'} = \frac{\rho^{i}}{1+Tr\left(\Delta E^{i}\right)} +
\frac{\Delta E^{i}}{1+Tr\left(\Delta E^{i}\right)},
\label{mapping}
\end{equation}
in which
\begin{equation}
\Delta E^{i} = Max\left\{0,
H^{i}_{R}-E^{i}\left(\rho^{S}\right)I^{i}\right\}, \label{deltaE}
\end{equation}
in which $Max$ means to get the bigger one between every element.
First, as a physicist usually does, let's show the fixed points
are NE. Denote the fixed points as $\rho_{eq}^{S} =
\prod_{i=1}^{N}\rho^{i}_{eq}$, then
\begin{equation}
\rho^{i}_{eq}Tr\left(\Delta E^{i}\right) = \Delta E^{i}.
\label{equilibrium}
\end{equation}
It's easy to know that
\begin{equation}
\Delta E^{i} = 0
\end{equation}
is one of the solutions. Let's suppose $\Delta E^{i} \neq 0$,
because every element is bigger than $0$, $Tr\left(\Delta
E^{i}\right)>0$. Therefor, from equ(\ref{equilibrium}), if the
diagonal element of $\rho^{i}_{eq,\mu\mu}>0$, $\Delta
E^{i}_{\mu\mu}>0$. Then from equ(\ref{deltaE}), the definition of
$\Delta E^{i}$, because it's bigger than zero,
\[
H^{i}_{R,\mu\mu} > E^{i}\left(\rho^{S}\right) =
\sum_{\nu}\rho^{i}_{eq,\nu\nu}H^{i}_{R,\nu\nu}, \forall
\rho^{i}_{eq,\mu\mu}>0.
\]
So the weighted average of $H^{i}_{R,\mu\mu}$,
\[
\sum_{\mu}\rho^{i}_{eq,\mu\mu}H^{i}_{R,\mu\mu} >
\sum_{\nu}\rho^{i}_{eq,\nu\nu}H^{i}_{R,\nu\nu}
\]
That's impossible. So $\Delta E^{i} = 0$ is the only solution.
Therefor, from equ(\ref{deltaE}),
\[
E^{i}\left(\rho^{S}_{eq}\right) \geq H^{i}_{R,\mu\mu}, \forall
\mu.
\]
And then, since it's bigger than every element of $H^{i}_{R}$,
it's bigger than any weighted average of them, so it's NE. All
players can't get more payoff by adjusting their own states
independently. Now we claim that this is a continuous and onto
mapping from system Hilbert space to itself. So it has fixed
points. This detailed proof is neglected here.

Now we have shown that the existence of Nash Equilibrium in our
new representation. The mapping defined here can be regarded as an
iteration starting from any arbitrary initial system state. But
does it converge onto the fixed points? Are the fixed points
stable? From NE and the proof of Nash Theorem, no thing we can say
about this question. The real experience in application of Game
Theory shows that sometime the NE is not stable. If they are
unstable, they are not very meaningful to be regarded as a
prediction of the game.

However, our representation here is quite similar with Quantum
Mechanics and Statistical Mechanics. We have state vector or
density matrix, and we have dynamical variables such as payoff
matrix and reduced payoff matrix, which look very like a
Hamiltonian. The only thing missing here is a dynamical equation,
which determines the evolution of state.

\section{Comparison between NE and PDE}
\label{compare}

In order to give an evolutionary equation, we recall Master
Equation for probability distribution function,
\begin{equation}
\frac{dp^{i}\left(x,t\right)}{dt} = \sum_{x^{'}}
w\left(x^{'}\rightarrow x\right)p^{i}\left(x^{'},t\right) -
\sum_{x^{'}} w\left(x\rightarrow
x^{'}\right)p^{i}\left(x,t\right),
\label{systemmaster}
\end{equation}
in which, we suppose the transition rate is
\[
w\left(x^{'}\rightarrow x\right) = \frac{e^{\beta
\left[E^{i}\left(x\right)-E^{i}\left(x^{'}\right)\right]}}{\sum_{y}e^{\beta
\left[E^{i}\left(y\right)-E^{i}\left(x^{'}\right)\right]}}=\frac{e^{\beta
E^{i}\left(x\right)}}{\sum_{y}e^{\beta E^{i}\left(y\right)}}.
\]
The Master Equation here is actually $N$ related equations,
because $E^{i}$ depends on $H^{i}_{R}$, which is determined by
other players' state. From background of Statistical Mechanics, we
know, if it's a single equation, or we say, $E^{i}$ is independent
on other players' state, the equilibrium state when
$\frac{dp^{i}\left(x,t\right)}{dt} = 0$ will be
\[
p^{i}\left(x,\infty\right) = \frac{e^{\beta
E^{i}\left(x\right)}}{\sum_{x^{'}}e^{\beta
E^{i}\left(x^{'}\right)}}.
\]
But, unfortunately, here all equations are related. So we make
another assumption that the time scale of a single Master Equation
is much smaller than the time scale of the related equations. In
physics, this means that we let the single equation evolute to
equilibrium first, then we feedback the equilibrium state into all
other equations and so on. Under such assumption, we will get
another $N$ related equations from equ(\ref{systemmaster}),
\[
p^{i}\left(x,t+1\right) = \frac{e^{\beta
E^{i}\left(x,t\right)}}{\sum_{x^{'}}e^{\beta
E^{i}\left(x^{'},t\right)}}.
\]
In density matrix notation,
\begin{equation}
\rho^{i}\left(t+1\right) = \frac{e^{\beta
H^{i}_{R}\left(t\right)}}{Tr^{i}(e^{\beta
H^{i}_{R}\left(t\right)})}, \label{iteration}
\end{equation}
very similar with the Boltzman distribution for canonical ensemble
in Statistical Mechanics. Now we have an evolutionary equation
although we don't know it can give some information about the game
solution or not. It's not on the basis of first principle,
however, as in Statistical Mechanics, since we wish it will give
reasonable game solution as equilibrium state, we name it
pseudo-dynamical equation and name the equilibrium state if
possible Pseudo-Dynamical Equilibrium (PDE).

In such evolution, first, we choose an initial state for every
player, at every step, start from player $i$, calculate
$H^{i}_{R}$, get the new state of player $i$ by
equ(\ref{iteration}), then feed it into other players' reduced
payoff matrix to get their new states. Repeat such step till some
fixed pattern if it's possible. The existence of such pattern is
not proved, and the specific order of choosing which player first
may affect such pattern. So the whole thing is still open, and
should be investigated further. Even the equation itself is
derived by two assumptions, first the Master Equation
(\ref{systemmaster}) and second the time scale assumption. The
intuitive meaning of such iteration equation is that every player
decides its own response to all other players according to the
possible payoff, but instead of choosing the best one uses a
distribution function, then other players repeat such iteration
and so on.

The privilege of our representation is that because we use payoff
matrix (a $(1,1)$-tensor) and reduced payoff matrix to calculate
the payoff value, when they are putted onto exponential function,
they will give a naturally defined density matrix of players'
states. Compared with the mapping of equ(\ref{mapping}), iteration
process equ(\ref{iteration}) can be approximately reexpressed as
\begin{equation}
\rho^{i}\left(t+1\right) =
\frac{\rho^{i}\left(t\right)}{1+Tr^{i}(e^{\beta
H^{i}_{R}\left(t\right)})} + \frac{e^{\beta
H^{i}_{R}\left(t\right)}}{1+Tr^{i}(e^{\beta
H^{i}_{R}\left(t\right)})},
\end{equation}
because usually $Tr^{i}(e^{\beta H^{i}\left(t\right)})\gg 1$. So
the difference between the mapping equ(\ref{mapping}) and our
iteration is that matrix $e^{\beta H^{i}_{R}\left(t\right)}$ is
used to replace matrix $\Delta E^{i}$. We wish such replacement
will not change the idea of NE so far that it still can give
information of game solution. On the other hand, besides the
similarity with the mapping, the iteration process here looks very
reasonable and comparable with the real game process. Everyone
chooses initial state first, then decides the best response
according to states of other players, and then iterates such
process.

However, whatever it looks like, the only test is whether it will
give reasonable game solution or not. A theoretical comparison
between PDE and NE is ongoing in our work plan, but here, as far
as the well-known specific games we tried, it gives quite good
results\cite{frame}. And in some cases, when unstable NE exists,
our iteration gives some pattern such as a jumping between some NE
states, in other cases, when stable NEs exist, it end at one point
of the NEs depending on initial state. It seems that even the
iteration process itself is meaningful. In such cases, stable NEs
can be regarded as the end results of the iteration process. It's
quite amazing, but still waiting for more exploration.

\section{Quantum Game and Quantum Nash Equilibrium}
\label{quantum}

From the proof of equivalence, equ(\ref{proof1}), we know that
because payoff matrix $H^{i}$ has only diagonal term, only
diagonal term of $\rho^{S}$ comes into the expression of payoff
value. And similarly, if $\rho^{S}$ has only the diagonal term,
only diagonal term of $H^{i}$ will be effective. This property
implies two things. The first, as we did for continuous strategy
games, a wave function can be used to replace distribution
function. Second, if in some games, a payoff function with
non-zero off-diagonal elements {\bf{and}} a non-zero off-diagonal
density matrix, are required, they will be totally {\bf{new
games}}. Effect of the first aspect will be discussed in the last
section ($\S\ref{remark}$). In this section, we try to find some
manipulative examples of the new games.

On the other hand, let's suppose players can make use of quantum
operator as strategy. We have a quantum object, every player
applies a quantum operator on it and then the payoff of every
player is determined by the end state of quantum object. In fact,
this game use the same idea of classical game, but with the
quantum object and quantum strategies. As in \cite{frame}, using
Quantum Penny Flip Game\cite{meyer} as example, the quantum object
is a spin, the quantum strategies are all unitary matrix acting on
the spin, and the payoff is determined by the end state of the
spin. Compared with classical games, in quantum game, base
strategies still can be defined, even quite natural. State space
of the quantum object has $Q$ base vectors
$\left\{\left|\mu\right>\right\}$, so the quantum operator has the
form
\begin{equation}
\hat{U} = \sum_{\mu\nu}U_{\mu\nu}\left|\mu\right>\left<\nu\right|.
\label{quantumstrategy}
\end{equation}
Therefor, $\left\{\left|\mu\right>\left<\nu\right|\right\}$ can be
regarded as $Q\times Q$ base vectors of the operator space. And
because the operator space is still a Hilbert space, we even can
define the inner product as
\begin{equation}
\left(\hat{U},\hat{V}\right) =
Tr\left(\hat{U}^{\dag}\hat{V}\right) \label{innerproduct}
\end{equation}
so that $\left\{\left|\Phi\right> =
\left|\mu\right>\left<\nu\right|\right\}$ will be a set of
orthogonal unit base vectors. Now we can denote the strategy state
of player $i$ by state vector or a density matrix in operator
Hilbert space,
\begin{equation}
\begin{array}{ccc}
\rho^{i} = \sum_{\Phi, \Psi} \rho^{i}_{\Phi\Psi}\left|\Phi\right>
\left<\Psi\right| & \mbox{or} & \rho^{i} = \left|U\right>
\left<U\right| = \sum_{\Phi,\Psi}
U_{\Phi}U^{*}_{\Psi}\left|\Phi\right> \left<\Psi\right|
\end{array}.
\end{equation}
In the direct product space of all single-player strategy spaces,
state of the whole $N$-player system is
\begin{equation}
\rho^{S} = \prod_{i=1}^{N}\rho^{i}, \label{systemstate}
\end{equation}
if the state of all players are independent, or in Game Theory
language, non-cooperative. According to Quantum Mechanics, the
payoff value should be
\begin{equation}
E^{i} = Tr\left(\rho^{S}H^{i}\right). \label{quantumpayoff}
\end{equation}
So a quantum game is defined as
\begin{equation}
\Gamma^{q} = \left(\prod_{i=1}^{N}\otimes S^{q}_{i},
\left\{H^{i}\right\}\right). \label{qgame}
\end{equation}

Some specific quantum games such as Quantum Penny Flip
Game\cite{meyer} and Quantum Prisoner's Dilemma\cite{jens} have
been reexpressed and studied in the new representation in
\cite{frame} and \cite{entangle}. Here we `theoretically' define a
manipulative general quantum game, and prove that it can be
described in our new representation. A real quantum game is
defined in quantum operator level, so all strategies are quantum
operators acting on a quantum object, whose initial state is
denoted as $\rho^{q}_{0}\in \mathbb{H}^{q}$. Here, because we use
density matrix, which can also be regarded as an operator, to
represent state of the quantum object, $\mathbb{H}^{q}$ denotes
both the state space of the quantum object and operator space on
it. Every player choose strategy $\hat{U}^{i} \in \mathbb{H}^{i}$,
which is an operator from subspace of $\mathbb{H}^{q}$ onto the
subspace, then the jointed operator acting on the quantum object
is
\begin{equation}
\hat{U} = \mathcal{L}\left(\hat{U}^{1}, \cdots,
\hat{U}^{N}\right),
\end{equation}
a linear mapping of $\hat{U}^{i}$ from direct product strategy
space $\mathbb{H} = \prod_{i}^{N}\otimes\mathbb{H}^{i}$ to
operator space of quantum object $\mathbb{H}^{q}$,
\begin{equation}
\mathcal{L}\left(\cdots, \alpha\cdot\hat{U}^{i},\cdots\right) =
\alpha\cdot\mathcal{L}\left(\cdots, \hat{U}^{i},\cdots\right),
\alpha \in \mathbb{C}, \forall i. \label{linear}
\end{equation}
Product and direct product are typical forms of such
mapping\cite{frame,entangle}. Then the end state of the quantum
object is
\begin{equation}
\rho^{q} = \hat{U}\rho^{q}_{0}\hat{U}^{\dag}.
\end{equation}
Payoff value of player $i$ is determined by
\begin{equation}
E^{i} = Tr\left(P^{i}\rho^{q}\right), \label{endpayoff}
\end{equation}
in which $P^{i}$ is a matrix in $\mathbb{H}^{q}$, named payoff
scale matrix, which gives the rule or scale to determine the
payoff. In Quantum Penny Flip Game, it's
\[
P^{1} = \left|U\right>\left<U\right| -
\left|D\right>\left<D\right| = -P^{2}.
\]
For player $1$, this means to assign $1$ to up state and $-1$ to
down state. The form in Quantum Prisoner's Dilemma Game has also
been given in \cite{entangle}. So a Quantum Game in operator
level, the language of Quantum Mechanics, is
\begin{equation}
\Gamma^{q,o} = \left(\mathbb{H}^{q}, \prod_{i=1}^{N}\otimes
\mathbb{H}^{i}, \mathcal{L}, \left\{P^{i}\right\}\right).
\label{qogame}
\end{equation}

Now as we did in classical game, we need to prove $\Gamma^{q,o}$
can be equivalently described by $\Gamma^{q}$. Before the detailed
proof, we want to point out the linear property of $\mathcal{L}$,
equ(\ref{linear}), is a very important condition. Even in
classical game, we can find the corresponding implied condition.
Starting from a pure strategy game, the only thing we know is the
elements of payoff tensor $G^{i}$, which can only give the payoff
for pure strategies, but we need to know the payoff value for
mixture strategies. Equ(\ref{oldpayoff}) uses mathematical
expectation to calculate it. It seems quite natural, but is it
really the only possible? The nonlinear behavior might be
possible. For example, consider the situation that a girl and a
boy work together. They have two optional jobs. When they use a
little time together, the interaction between them is weak, the
efficiency is low; when they spend more time together, they know
each other better, so the efficiency is higher; when they spend
too much time together, they will find more shortcoming of each
other, or otherwise they will flirt with each other, anyway, the
efficiency will be lower again. This is a truly non-linear
behavior. Our current Game Theory could never describe this
phenomenon. So this is the implied linear condition of classical
game,
\begin{equation}
E^{i}\left(\cdots, \alpha \cdot \rho^{i}_{a} + \beta \cdot
\rho^{i}_{b}, \cdots \right) = \alpha \cdot E^{i}\left(\cdots,
\rho^{i}_{a}, \cdots \right) + \beta \cdot E^{i}\left(\cdots,
\rho^{i}_{b}, \cdots \right), \forall
\alpha,\beta\in\left[0,1\right].
\end{equation}
This will require $\left\{G^{i}\right\}$ are linear mappings,
$\left(0,N\right)$-tensors. Now in our quantum game, this
condition implies linear property of mapping $\mathcal{L}$ and
equ(\ref{endpayoff}), the trace operator. And in the abstract form
$\Gamma^{q}$, this condition is automatically fulfilled when
$\left\{H^{i}\right\}$ are $\left(1,1\right)$-tensors. Now we
prove the equivalence.

\noindent {\bf{Theorem III}} $\Gamma^{q,o}$ is equivalent with
$\Gamma^{q}$. This is to say for all players with arbitrary
strategies, the two representations give the same payoff for every
player.

\noindent {\bf{Proof}} For player $i$'s operator space, choose a
set of base vectors according to equ(\ref{innerproduct}), the
inner product definition, and denote them as
$\left\{\left|s\right>^{i}\right\}$ and
$\left\{\hat{s}^{i}\right\}$, which, for the quantum object, are
operators such as $\left\{\left|\mu\left>
\right<\nu\right|\right\}$, but for the players, are some base
strategies. Suppose, player $i$ choose strategy $\hat{U}^{i}$,
which can be expanded as
\[
\hat{U}^{i} = \sum_{s}U^{i}_{s}\left|s\right>^{i} =
\sum_{s}U^{i}_{s}\hat{s}^{i}.
\]
Density matrix form of this player's state is
\[
\rho^{i} =
\sum_{\phi^{i},\psi^{i}}U^{i}_{\phi}\left(U^{i}_{\psi}\right)^{*}\left|\phi\right>^{i}\left<\psi\right|^{i}.
\]
System state density matrix is
\[
\rho^{S} =
\prod_{i}^{N}\sum_{\phi^{i},\psi^{i}}U^{i}_{\phi}\left(U^{i}_{\psi}\right)^{*}\left|\phi\right>^{i}\left<\psi\right|^{i}.
\]
Define every elements payoff matrix in this representation as
\begin{equation}
\left<\phi^{1}\cdots\phi^{N}\right|H^{i}\left|\psi^{1}\cdots\psi^{N}\right>
= H^{i}_{SS^{'}} = Tr\left(P^{i}
\mathcal{L}\left(\hat{\phi}^{1},\cdots,\hat{\phi}^{N}\right)\rho^{q}_{0}\mathcal{L}^{\dag}\left(
\hat{\psi}^{1}, \cdots, \hat{\psi}^{N}\right)\right).
\end{equation}
Payoff from equ(\ref{endpayoff}) is
\[
\begin{array}{ccc}
E^{i}  & = & Tr\left(P^{i} \mathcal{L}\left(\cdots,
\sum_{\phi^{j}}U^{j}_{\phi}\hat{\phi}^{j},
\cdots\right)\rho^{q}_{0}\mathcal{L}^{\dag}\left(\cdots,
\sum_{\psi^{j}}U^{j}_{\psi}\hat{\psi}^{j}, \cdots\right)\right)
\\
\\ & = & Tr\left(P^{i} \sum_{\phi^{j}}U^{1}_{\phi}\cdots U^{N}_{\phi}\mathcal{L}\left(\cdots,
\hat{\phi}^{j},
\cdots\right)\rho^{q}_{0}\sum_{\psi^{j}}\left(U^{1}_{\psi}\right)^{*}\cdots\left(U^{N}_{\psi}\right)^{*}\mathcal{L}^{\dag}\left(\cdots,
\hat{\psi}^{j}, \cdots\right)\right)
\\
\\ & = & \sum_{\phi^{j}}U^{1}_{\phi}\cdots U^{N}_{\phi}\sum_{\psi^{j}}\left(U^{1}_{\psi}\right)^{*}\cdots\left(U^{N}_{\psi}\right)^{*}Tr\left(P^{i} \mathcal{L}\left(\cdots,
\hat{\phi}^{j},
\cdots\right)\rho^{q}_{0}\mathcal{L}^{\dag}\left(\cdots,
\hat{\psi}^{j}, \cdots\right)\right)
\\ \\ & = & \sum_{\phi^{j}}U^{1}_{\phi}\cdots U^{N}_{\phi}\sum_{\psi^{j}}\left(U^{1}_{\psi}\right)^{*}\cdots\left(U^{N}_{\psi}\right)^{*}\left<\phi^{1}\cdots\phi^{N}\right|H^{i}\left|\psi^{1}\cdots\psi^{N}\right>
\\
\\ & = & Tr\left(\rho^{S}H^{i}\right)
\end{array}.
\]
So for all quantum pure strategy, those two forms give the same
payoff. We require $P^{i}$ is hermitian. It's a payoff scale
matrix, which assigns values to every state, therefor, in a
specific set of base vectors, $P^{i}$ should have only diagonal
terms. So in this representation, $P^{i}_{\mu\nu} = 0 =
\left(P^{i}_{\nu\mu}\right)^{*}$, then $P^{i}=
\left(P^{i}\right)^{\dag}$ generally. Therefor, $H^{i}$ is also
hermitian. Further more, in density matrix form, not only pure
quantum strategies, but also quantum mixture strategies are
allowed to be used by quantum players. And still the payoff are
given by equ(\ref{quantumpayoff}). This means, not only unitary
operator, but also mixture of unitary operators, can be a strategy
of quantum player. Whether such strategy is applicable or not will
depend on further application and realization of the idea of
quantum game.

Now classical game and quantum game have the same forms except
non-zero off-diagonal terms in quantum payoff matrix. Sometimes,
quantum game has classical sub-game, in which a set of special
base vectors (strategies) can be found that the corresponding
sub-matrix $H^{i,c}$ of $H^{i}$ is diagonal. An example of this is
Quantum Penny Flip Game and Classical Penny Flip Game\cite{frame},
in which $N^{c}, F^{c}$ are classical base vectors while the
quantum base vectors include other two base strategies $N^{q},
F^{q}$. In that situation, we say the quantum game has classical
limit. So there are several different strategy spaces, classical
pure strategy space, classical mixture strategy space, quantum
pure strategy space and quantum mixture strategy space. Even a
larger space can be took into our consideration, which destroys
equ(\ref{systemstate}), the direct product relation between system
state and single-player state. We name it entangled strategy
space, which includes vectors in system space not only the
direct-product states. And we name the game with diagonal payoff
matrix in classical mixture strategy space as Classical Game (CG);
the game with diagonal payoff matrix in quantum pure strategy
space as Quantized Classical Game (QCG), which are equivalent with
CG; the game with general payoff matrix in quantum pure strategy
space as Pure-strategy Quantum Game (PQG); the game with general
payoff matrix in quantum mixture strategy space as Quantum Game
(QG); the game with general payoff matrix in entangled quantum
strategy space as Entangled Quantum Game (EQG), the game with
diagonal payoff matrix in entangled classical strategy space as
Entangled Classical Game (ECG). So
\[
\left\{
\begin{array}{c}
CG = QCG \subseteq ECG
\\
PQG \subseteq QG \subseteq EQG,
\end{array}\right.
\]

Now our question is what kind of equilibrium states exist in which
strategy space. We know Nash Theorem is valid in classical mixture
strategy space. Will a Nash-like equilibrium state exist in
quantum game, and in which strategy space? A General Classical
Nash Equilibrium (GCNE) and General Quantum Nash Equilibrium
(GQNE) is defined as
\begin{equation}
E^{i}\left(\rho^{S}_{eq}\right)\ge
E^{i}\left(Tr^{i}\left(\rho^{S}_{eq}\right)\cdot\rho^{i}\right),
\forall \rho^{i}, \forall i.
\end{equation}
When $\rho^{S} = \prod_{i}^{N}\rho^{i}$, this definition will
become equ(\ref{oldne}), the traditional definition of NE, but for
quantum game, we name it Quantum Nash Equilibrium (QNE).

\noindent {\bf{Proposition}} NE exists in CG, GCNE exists in ECG,
QNE exists in QG and GQNE exists in EQG.

\noindent {\bf{Proof}} A general proof is still open. The first
part is Nash Theorem. For the third part, for a special class of
quantum game, in which $\left[H^{i},H^{j}\right] = 0, \forall
i,j$, we can prove the existence of QNE. First, find the common
eigenvectors set of $\left\{H^{i}\right\}$, and then in this new
representation, all payoff matrix are diagonal. So the quantum
game looks similar with classical game, but in another set of base
vectors. Because NE exists in the new classical game, QNE exists
in the original quantum game. So for such games, QNE exists. For
the second and the fourth part, for another special class of
quantum game, in which all $\left\{H^{i}\right\}$ have a common
eigenvector $\rho^{S}_{M}$ with maximum eigenvalue, so that
\begin{equation}
E^{i}\left(\rho^{S}_{M}\right)\ge E^{i}\left(\rho^{S}\right),
\forall \rho^{S}, \forall i, \label{special}
\end{equation}
$\rho^{S}_{M}$ will be GCNE or GQNE. Since it's a vector in system
space, it's possible that it is not a direct product of single
player state. Let's keep fingers crossed for a general proof in
the near future.

\section{Conclusion and discussion}
\label{remark}

We have seen that our Quantum Game and Entangled Quantum Game are
truly independent thing, which is impossible to be put into the
old framework of $G^{i}$ in $\Gamma^{c}$. In fact, elements of
$G^{i}$ are only the diagonal part of $H^{i}$. This is partially
an answer to the second question in \cite{enk}, but still not a
confirmative answer. The applicability value of this Quantum Game
can only been shown through a real quantum game, in which players
make use of quantum operators acting on a quantum object. However,
most examples we have now are toy games, not from real experiments
in quantum world. In the future, we will try to propose one real
game from the world of quantum computation or quantum
communication, etc.

Now, we turn to discuss the first question in \cite{enk}, if our
new approach is helpful to solve the Classical Game. Since $G^{i}$
is enough for classical game, theoretically, this new approach
brings nothing new into Classical Game. So how about the technical
level? Will the new approach be helpful to calculation of NE in
Classical Game? We have shown in section $\S$\ref{continue}, a
wave vector and its density matrix $\rho^{i} =
\sum_{\mu\nu}\left(\phi^{i}_{\nu}\right)^{*}\phi^{i}_{\mu}\left|\mu\right>\left<\nu\right|$
can equivalently replace $\vec{p}^{i} =
(\cdots,p^{i}_{\mu},\cdots)^{T}$. In fact, the former provides
much redundant information through the off-diagonal terms, as we
know, in classical game, only the diagonal terms
$p^{i}_{\mu}=\left(\phi^{i}_{\mu}\right)^{*}\phi^{i}_{\mu}$ effect
the payoff. From theoretical viewpoint, it's waste of time,
however, from calculation viewpoint, this means we can use quantum
state to calculate problems in Classical Game. Because a single
quantum state includes as much as variables in a very complex
combination classical state, this makes it possible to improve
calculation of NE in Classical Game. Further more, if an evolution
equation of wave vector, instead of our pseudo-dynamical equation
of density matrix, can be found, next time, when we need to
calculate classical NE, the only thing need to do is to choose an
arbitrary initial state of a quantum system and let it evolute
according to the equation, and then, the end state will be the
answer. As in Quantum Mechanics, Schr\"{o}dinger Equation of wave
vector is equivalent with Liouville Equation of density matrix,
all linear equation of density matrix can be transferred into
linear equation of wave vector. Although our current
pseudo-dynamical equation are not linear, it is not impossible to
find another better linear equation to describe the evolution
process.

Besides the technical level of NE calculation, the new approach
opens an exited way to deal with Evolutionary Game and Cooperative
Game . From the application experience in \cite{frame}, not only
the end state, but also the pseudo-dynamical process seems
meaningful in Game Theory. And in the new representation, the
system state of non-cooperative players is a direct product
density matrix, while a general density matrix in system state
space includes some correlation between players, can be naturally
used to discuss cooperative game. Evolutionary Game and
Cooperative Game are another two important topics. In traditional
Game Theory, the framework of evolutionary game is far from
elegant. It will be a great progress if this new representation
can be easily applied onto those two aspects. For example, let's
suppose a GCNE/GQNE, or specially the $\rho^{M}$ in
equ(\ref{special}) has been found for a game. It's not a
direct-product state, so that
\begin{equation}
\rho^{S} \neq \prod_{i}^{N}Tr_{-i}\left(\rho^{S}\right).
\end{equation}
But it's still probable in some sectors of players $N =
\bigcup^{s}_{j=1}N^{j}$, in which $N$ is the set of players and
$N^{j}$ is a subset of $N$, that
\begin{equation}
\rho^{S} = \prod_{j}^{s}\rho^{j},
\end{equation}
although when $\left|N^{j}\right|>1$, $\rho^{j} \neq
\prod_{i}^{\left|N^{j}\right|}Tr_{-i}\left(\rho^{j}\right)$. So
the subsets $\left\{N^{j}\right\}$ can be regarded as player
groups. Therefor, if such groups can be naturally derived from ECG
and EQG, our new representation will be a way being applicable to
Cooperative Game\cite{gamecourse}.

For the applicability value of our quantum game, another factor
has to be took into consideration. Not all vectors in quantum pure
strategy space are unitary operators. If we require all physical
operators are unitary, only part of the space can give real
applicable strategies, and even worse, they might not be a
subspace. This will exclude some equilibrium solutions, even when
QNE or GQNE does exist in the whole space of quantum strategy
space or entangled quantum strategy space. In that situation, it's
not a confirmative result even we prove the existence of NE.
Fortunately, in a $2\times 2$ quantum operators space, the unitary
operators still expand a subspace. A general $2\times 2$ quantum
operator is
\begin{equation}
\hat{U} = \xi\cdot\hat{I} + x \cdot\hat{\sigma}_{x} + y
\cdot\hat{\sigma}_{y} + z \cdot\hat{\sigma}_{z}, \xi,x,y,z \in
\mathbb{C},
\end{equation}
while a unitary $2\times 2$ quantum operator is
\begin{equation}
\hat{U} = \xi\cdot\hat{I} + ix \cdot\hat{\sigma}_{x} + iy
\cdot\hat{\sigma}_{y} + iz \cdot\hat{\sigma}_{z}, \xi,x,y,z \in
\mathbb{R}.
\end{equation}
So we can still ask the question that if QNE and GQNE exist in the
subspace. But we only have such good conclusion for $2\times 2$
quantum operator space, we don't know it's a general result or
not. However, even if it's not a general conclusion so that we
have to discuss equilibrium state in the whole space including
non-unitary operators, maybe someday, we can make use of
non-unitary operators as strategy so it's even better to consider
questions in the whole space. All such open problems depend on the
application and realization of real quantum games.

\section{Acknowledgement}
The authors want to thank Dr. Qiang Yuan, Shouyong Pei and Zengru
Di for their advices during the revision of this paper. This work
is partial supported by China NSF 70371072 and 70371073.


\begin{thebibliography}{9}
\bibitem{frame} Jinshan Wu, A new mathematical representation of Game
Theory, arXiv:quant-ph/0404159.

\bibitem{enk} S. J. van Enk and R. Pike, Classical rules in quantum
games, Phys. Rev. A {\bf{66}}(2002), 024306.

\bibitem{meyer}D.A. Meyer, Quantum Strategies, Phys. Rev. Lett. {\bf{82}}(1999),
1052.

\bibitem{jens} J. Eisert, M. Wilkens, and M. Lewenstein, Quantum Games and
Quantum Strategies, Phys. Rev. Lett, {\bf{83}}(1999), 3077.

\bibitem{entangle}Jinshan Wu, An artificial game with entangled equilibrium
state, arXiv:quant-ph/0405003.

\bibitem{gamecourse}M. J. Osborne and A. Rubinstein, {\it A course in
Game Theory}, MIT Press, 1994.
\end{thebibliography}
\end{document}